%% file: main.tex
\documentclass[sigconf]{acmart}
\setcopyright{none}
\acmConference[Arxiv pre-print]{}{September}{2024}

\usepackage{threeparttable}
\usepackage[utf8]{inputenc}
\usepackage{graphicx}
\usepackage{subcaption}
\usepackage{fancyvrb}
\usepackage{multirow}
\usepackage{makecell}
\usepackage{array, multirow}
\usepackage{arydshln}
\newcommand{\ignore}[1]{}
\usepackage{balance}
\usepackage{xcolor}
\usepackage[ruled,vlined]{algorithm2e}
\usepackage{booktabs}
\usepackage{listings}

\usepackage{pifont}
\newcolumntype{P}[1]{>{\centering\arraybackslash}p{#1}}

\usepackage{array}
\newcolumntype{H}{@{}>{\setbox0=\hbox\bgroup}c<{\egroup}}

\usepackage{tikz}

\begin{document}

\title{RPKI: Not Perfect But Good Enough}

\author{Haya Schulmann}
\affiliation{\small{ATHENE \& Goethe-Universität Frankfurt\country{Germany}}}

\author{Niklas Vogel}
\affiliation{\small{ATHENE \& Goethe-Universität Frankfurt\country{Germany}}}

\author{Michael Waidner}
\affiliation{\small{ATHENE \& TU Darmstadt\\\country{Germany}}}

\input{abstract}

\maketitle

\input{introduction}
\input{background}

\input{vuln}
\input{attacks}

\input{conclusion}
\balance

\bibliographystyle{ACM-Reference-Format}
\bibliography{ref}

\end{document}

%% file: abstract.tex
\begin{abstract}
The Resource Public Key Infrastructure (RPKI) protocol was standardized to add cryptographic security to Internet routing. With over 50\% of Internet resources protected with RPKI today, the protocol already impacts significant parts of Internet traffic \cite{nistmonitor}.
In addition to its growing adoption, there is also increasing political interest in RPKI. The White House indicated in its \textit{Roadmap to Enhance Internet Routing Security}, on 4 September 2024, that RPKI is a mature and readily available technology for securing inter-domain routing. The Roadmap attributes the main obstacles towards wide adoption of RPKI to a lack of understanding, lack of prioritization, and administrative barriers. 

This work presents the first comprehensive study of the maturity of RPKI as a viable production-grade technology. We find that current RPKI implementations still lack production-grade resilience and are plagued by software vulnerabilities, inconsistent specifications, and operational challenges, raising significant security concerns. The deployments lack experience with full-fledged strict RPKI-validation in production environments and operate in fail-open test mode. 
We provide recommendations to improve RPKI resilience and guide stakeholders in securing their deployments against emerging threats. 

The numerous issues we have discovered with the current RPKI specifications and implementations inevitably lead to the question: Is RPKI sufficiently stable to align with the expectations outlined in the White House roadmap? Certainly, it is not perfect, but is it good enough? The answer, as we will explore, varies depending on one's viewpoint.

\end{abstract}

%% file: introduction.tex
\section{Introduction}
The Border Gateway Protocol (BGP) underlies all modern Internet communication by enabling the exchange of IP routing information between Autonomous Systems (AS). Despite its criticality, BGP is insecure by design, and to this day, attacks on BGP are frequent and detrimental \cite{DBLP:journals/cacm/SunABVRCM21,klayswap, cloudfarehijack}. The Resource Public Key Infrastructure (RPKI) was standardized to add security to BGP through cryptographic attestations called Route Origin Authorizations (ROAs) \cite{rpkirfc}. The publication of RFC6811 introduced Routing Origin Validation (ROV), enabling ASes to use ROAs to verify if a given BGP announcement was originated by the legitimate holder of an Internet resource. This ability to validate legitimate ownership guides routers in their decisions and prevents attacks on BGP.

{\bf Chronicles of RPKI.} After its introduction \cite{nanog:history,nanog:history2}, RPKI was considered an experimental technology and its adoption was sparse. After 2010, RPKI started transitioning from a theoretical concept to practical implementations. Early software development was often led by research institutions and individual engineers working on open-source RPKI tools. In 2011 RIPE NCC launched one of the first tools for validating ROAs, called the RPKI Validator.\footnote{\url{https://labs.ripe.net/author/nathalie_nathalie/lifecycle-of-the-ripe-ncc-rpki-validator/}} During the early experimental phase in 2013, RPKI was mostly experimented with by some institutions, such as CAIDA and RIPE NCC, and a number of ISPs and carriers. While interest in RPKI was growing, the technology had only been deployed in experimental or partial configurations by around a dozen ISPs globally by that point, mainly in testbeds and pilot environments to gauge the impact on routing without affecting real traffic%
\footnote{\url{https://archive.nanog.org/sites/default/files/04-Murphy-StLouis.pdf}}. RIPE NCC and ARIN were among the first to provide RPKI services for operators to create and publish ROAs. These efforts focused on building the infrastructure for certificate issuance and ROA validation. Following 2016, more open source relying party software packages were created, offering operators more variety in adopting RPKI. Nevertheless, the RPKI implementations were mostly research prototypes rather than stable production software.
In 2018, major Internet stakeholders, such as the Regional Internet Registries (RIRs), began promoting RPKI adoption and large networks, service providers, and Content Delivery Networks (CDNs) started testing RPKI. In a surprising step Cloudflare announced full support for RPKI validation in 2018, marking a significant milestone as one of the first large public networks to enable ROV in a production environment%
\footnote{\url{https://blog.cloudflare.com/rpki-details/}}. In 2020, the Mutually Agreed Norms for Routing Security (MANRS) initiative pushed for widespread ROV adoption, emphasizing routing security and promoting best practices. RIPE NCC also advocated for RPKI deployment and ROV enforcement. The following year, more large ISPs and network providers, including Google and Amazon, began implementing ROV, helping to drive global adoption and encouraging other networks to follow suit. A notable increase in ROAs publication was observed globally in 2022. This allowed more networks to validate the authenticity of BGP announcements, even though full ROV enforcement across all networks was not yet in place. These milestones indicate the significance that RPKI is gradually gaining in the Internet.
Starting off as an experimental technology, RPKI became a central component in the Internet and already affects a large fraction of the networks. Today, more than 50\% of announced prefixes are covered with ROAs, and about 25\% of networks enforce ROV \cite{nistmonitor, hlavacek2023keep, li2023rovista}. %But did RPKI actually transition from experimental to a mature production-grade technology?

{\bf Growing political significance.} 
The US was one of the first countries to recognize the threat that vulnerable Internet routing introduces to national security and the necessity of RPKI to secure their inter-domain BGP routing. It thus listed routing security as one of the main items on its cybersecurity agenda\footnote{\url{https://www.whitehouse.gov/wp-content/uploads/2023/03/National-Cybersecurity-Strategy-2023.pdf}} in March 2023. Shortly after, in September 2024, the White House issued a strategic roadmap. The roadmap establishes a plan of action and promotes the use of RPKI in all US networks with the goal of improving routing security against attacks \cite{whreport}. The roadmap complements a Notice of Proposed Rulemaking (NPRM) of the United States Federal Communications Commission (FCC), published a few months earlier in May 2024. Both documents provide strategic and policy frameworks as well as technical steps, with focus on compliance and operation. The roadmap of the White House identifies RPKI as a mature, ready-to-implement technology to mitigate vulnerabilities in BGP, and recommends to deploy it on all networks: 

``{\em The roadmap released today advocates for the adoption of Resource Public Key Infrastructure (RPKI) as a mature, ready-to-implement approach to mitigate vulnerabilities in BGP.}''

The roadmap considers the RPKI technology to be readily available and traces the challenges to RPKI adoption to three factors. First, decision makers lack a thorough understanding of the Internet routing security risks. Second, network operators do not prioritize and do not have resources to deploy new BGP security mechanisms. Third, organizations encounter administrative barriers with RIRs during adoption. The conclusion is that these challenges contribute to a reluctance to prioritize routing security. 

{\bf Recommendations for extending the roadmap.} In this work we develop an analysis of the gaps that need to be bridged to facilitate wide adoption of RPKI. Our analysis covers standard specifications, software packages, operation and deployment. We propose to enhance the roadmap to also include these concrete items. 
 We use the insights and observations from our research to derive recommendations for the different stakeholders in RPKI and routing security to provide a path forward for securing RPKI: 
 We recommend to collect experience with operating RPKI technologies in strict validation mode and recommend to transition networks to production-mode RPKI validation. We recommend to refine the RPKI standards, to remove conflicting or under-specified requirements, making clear recommendations for developers and operators of RPKI. We recommend to invest in developing RPKI-specific automated tools for software developers and operators and in prioritizing and automating patch-management. We suggest updates to the guidelines and strategies to reflect the full threat landscape of RPKI. The current guidelines in the roadmap exclusively focus on RPKI operation under benign conditions. We recommend considering resilience of RPKI under malicious attacks against its components. Finally, we recommend to consider the increase in the attack surface associated with networks deploying RPKI, by exploiting not only benign bugs but also intentional backdoors.

Improving routing security is a global and complex effort. We encourage that more countries contribute to the goal of securing Internet routing with RPKI, and hope that our recommendations will inform and guide their efforts.

{\bf Organization.} We review RPKI in Section \ref{sc:overview}. In Sections \ref{sc:spec} - \ref{sc:surface} we discuss various factors in RPKI resilience and security that need to be addressed for RPKI to become a mature technology. We discuss the role of this research in Section \ref{sc:role} and conclude in Section \ref{sc:conclusions}.

%% file: background.tex
\section{Overview of BGP Security}\label{sc:overview}
BGP lacks cryptographic authentication of announcements, which enables a range of routing attacks. 

{\bf Routing attacks.} In origin-hijack attacks, Autonomous Systems, ASes, can forge the BGP origin and claim to own arbitrary IP prefixes to hijack parts of the traffic to the hijacked prefix \cite{cho2019bgp}. This attack can be made more severe by announcing a sub-prefix of the attacked prefix, which will lead to all traffic being sent to the attacker. Running an origin hijack enables a range of other attacks, including DoS through blackholing the traffic to the hijacked prefix, and eavesdropping or other man-in-the-middle attacks \cite{butler2009survey}. 
In addition to the origin, BGP announcements also contain the path that a given announcement took, which corresponds to the path that traffic will take to the origin prefix. Instead of claiming to be the origin of a prefix, the attacker can also pretend to be on the path to the legitimate origin to achieve a path hijack. This allows similar attacks to origin hijacks, but is generally more difficult to detect, as the origin of the message is valid. 
Another substantial issue in BGP is route leaks. BGP dictates propagation paths of announcements should align with the business relationship of systems. For example, a customer should never provide transit services for its provider, since it is not paid by the provider to do so. The strict propagation rules ensure stability of the global routing system \cite{gao2000stable}. Paths that violate propagation rules are considered route leaks. BGP does not provide any mechanisms to prevent route leaks, and they can generally happen due to misconfigurations or, in rare cases, malicious attacks. The impact of route leaks can range from DoS caused by a system getting overloaded with traffic to instabilities in routing, with routes flapping between available and unavailable. For example, a recent route leake made Cloudflare DNS services unavailable for some users.\footnote{\url{https://blog.cloudflare.com/cloudflare-1111-incident-on-june-27-2024/}} A route leak might also allow eavesdropping, e.g., if a customer provides transit to its provider to access the traffic.

\begin{figure}[t!]
    \centering
    \includegraphics[width=0.8\columnwidth]{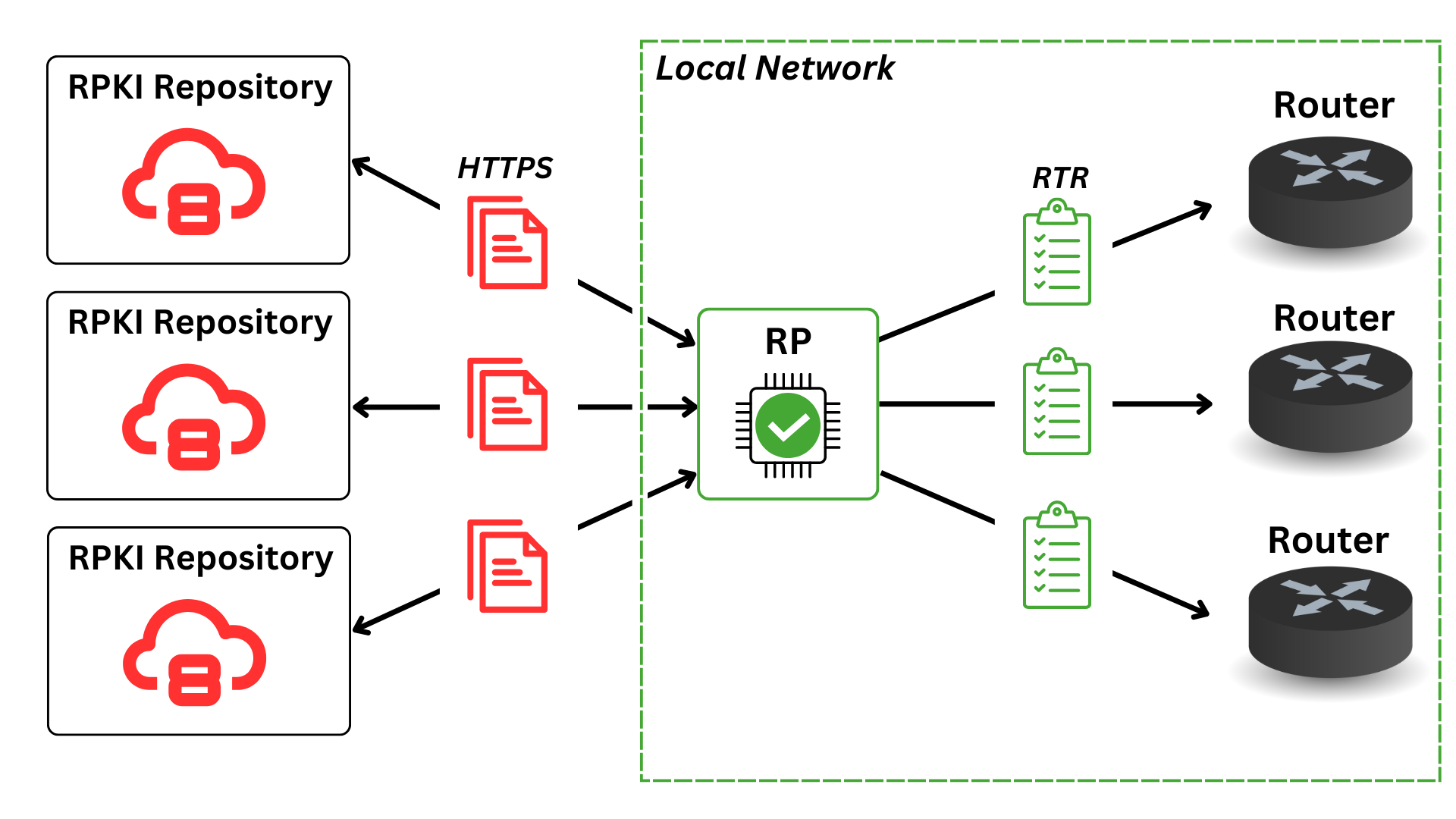}
    \vspace{-10pt}
    \caption{Overview RPKI.}
    \vspace{-15pt}
    \label{fig:rpki}
\end{figure}

\textbf{RPKI overview.} RPKI was designed to mitigate attacks BGP by adding cryptographic security to the insecure BGP protocol \cite{lepinski2012rfc}. RPKI provides an architecture to distribute verifiable BGP information globally. An overview of RPKI from the perspective of a system operator is given in Figure \ref{fig:rpki}. The red color illustrates untrusted RPKI components controlled by third parties, while the green color marks components trusted by the system and running within the local network.

\textbf{RPKI repository.}
All BGP information within the RPKI is stored inside distributed repository servers. RPKI offers two different hosting models: In hosted mode, participating systems create their RPKI objects within a repository operated by one of the five RIRs, while delegated mode allows systems to host their own repository servers, allowing full control over created objects. Both modes offer benefits, with hosted mode significantly reducing setup effort for systems, while delegated mode gives much more control over issued objects. Further, delegated mode improves decentralization of the architecture. Independent of the hosting models, systems accessing RPKI repositories should treat them as untrusted entities, and validity of contained objects should only be assumed through cryptographic validation (RFC6481).

\textbf{RPKI objects.}
RPKI offers a range of objects that systems can upload to RPKI repositories to provide data for BGP routing. 
Most prominently, ROAs contain information about the valid BGP origin of a set of IP prefixes, i.e. systems can use ROAs to verify if the origin of a given BGP announcement has been authorized by the owner of the announced prefix, protecting against origin hijacks \cite{roarfc}. 
Additional objects include BGPsec certificates \cite{lepinski2017bgpsec} that contain a public key to verify a signed BGP path, preventing path hijacks, and AS Provider Authorizations (ASPAs) that verify the valid provider set of a system to prevent route leaks \cite{asparfc}. Further, RPKI defines objects that ensure integrity of other objects and provide attestation of resource ownership, e.g. to allow verification if a given system is authorized to issue ROAs for a set of IP prefixes. This ownership of resources is verified over X.509 certificates that bind a cryptographic key to a set of IP resources, and are signed by the parent that issued the IP resources \cite{x509rfc}.

\textbf{Relying Party.}
To reduce computational load and implementation complexity, BGP routers do not directly interact with RPKI repositories to download and validate the RPKI objects. Instead, systems install a Relying Party (RP) client as a middleware between the RPKI repository system and their routers. The RP handles all interactions with the repositories, it downloads all objects from the available RPKI repositories, validates their integrity and cryptographic signatures, and compiles a list of all RPKI data for the routers. RPs recursively iterate all RPKI repositories on the Internet, using hardcoded URIs of the five hosted repositories for bootstrapping before iteratively querying the objects from all delegated repositories. Thus, each live RPKI repository will be regularly contacted by all globally running RP clients to download the contained RPKI objects. Since the RP is responsible for validation and inherently trusted, it should be installed within a trusted environment, like the local network. 

\textbf{BGP Router.}
All BGP routers in a system that want to utilize RPKI information regularly poll the RP to download the validated RPKI data over the RPKI to Router (RTR) protocol. Routers thus heavily depend on the availability of the relying party; \textit{if the relying party is unavailable, routers will eventually downgrade to regular BGP}  \cite{bush2014rfc}. Thus, RPs need to ensure availability to routers. Further, routers do not conduct any cryptographic validations of the data provided to them by their RP, the RP is fully trusted. Due to the availability and trust requirements of RPs, it is generally recommended to place RPs close to the routers, preventing disruptions of the connection and reducing the threat of manipulation of the communication \cite{bush2017resource}. After downloading the RPKI information, routers may use the data in their routing decisions. For example, they should use the ROA data to validate the origin in received BGP announcements, and discard any announcements that conflict with a ROA. Similarly, the routers may use ASPAs to reject announcements that constitute a route leak and BGPsec keys to validate received BGPsec paths to protect against path hijacks.

{\bf Maturity of RPKI technology.} 
The White House roadmap uses the term "mature" with respect to RPKI in a rather informal way. But actually, there are generally accepted models to characterize the maturity of a technology. The most common one is the Technology Readiness Levels (TRL) defined by NASA in the 1970s and adapted to assess the maturity of the IT technology as a Capability Maturity Model Integration (CMMI) \cite{mankins1995technology,tyson2003interpreting}. They range from TRL 1, basic principles are conceptualized, to TRL 9, technology is mature and ready for full-scale use in the real world. Clearly, RPKI is above TRL 1, but where is it on this scale? We will come back to this question, after having reviewed the resilience and stability of various aspects of the RPKI technology, such as the RPKI specifications, software packages and patch management, RPKI operation, and RPKI deployments.

%% file: vuln.tex
\section{Maturity of RPKI Specification}\label{sc:spec}

The Internet Engineering Taskforce (IETF) defined about 40 RPKI-related RFCs, which are generally complementary, so that each document addresses a specific aspect of the RPKI ecosystem. Although the multiple standards aim to address missing or vague details and guide the developers in implementing RPKI, their large number and often conflicting requirements increase the complexity and hence the risk of bugs and vulnerabilities. We next explain potential problems that arise from conflicting requirements or under-specification and provide concrete examples from our own evaluations and Internet measurements.

\subsection{Conflicting Requirements}
Conflicting requirements can result in inconsistent operational choices made by the networks that adopt the updated RFCs and those that continue to follow earlier ones. Discrepancies in the operational choices can lead to routing instabilities, rejected routes, and even security vulnerabilities. In our analysis we find many conflicting requirements in the validation rules, particularly those around ROA validity, certificate expiration, and error handling. The conflicting validation rules create small windows where a network could unintentionally accept hijacked prefixes. By adhering to older or more permissive requirements or being too flexible in handling invalid or expired objects, networks might allow unauthorized route announcements to propagate, exposing them to prefix hijacks.

{\bf Discrepancies in Filtering Logic.} For instance, RFC6811 introduces strict origin validation with ROAs. Further, RFC8416 introduces an additional mechanism (SLURM) to overwrite RPKI validity for certain routes. Using SLURM, a network operator may fix known issues with invalid routes, e.g., manually overwrite invalidity of multi-homed announcements only covered by a single ROA. Such cases are still prevalent to-date \cite{nistmonitor}. Systems not using SLURM to fix these problems will conclude different validity status of routes compared to systems with manually configured rules, leading to inconsistencies in the global routing table and potential route hijacking risks for networks not manually configuring required SLURM rules to fix such issues with missing ROAs. Vice-versa, the ability to configure arbitrary SLURM rules circumventing RPKI data also increases the potential for misconfiguration in systems, and can thus also open systems up to hijacks.

{\bf Discrepancies Which Routes to Filter.} A further example on discrepancies in validation is inherent in RFC6811, which is clarified by an update RFC8481. The phrasing on which received routes to validate sets a strong focus on BGP Update messages, leaving open if other related messages, like iBGP, should be validated in the first place. Further, it leaves ambiguity how to validate other messages, like which AS number should be used for validation. RFC8481 clarifies potential misunderstandings, stating \textit{all} routes should be validated and which ASN to use for validation in which cases. While operators that read the clarification will apply filtering as intended by the initial RFC, the large amount of published RPKI RFC makes it likely many operators do not keep track of new published RFCs that update existing recommendations and the diverse interpretations of the initial RFC will thus lead to inconsistent routing behavior between systems. 

\subsection{Vague and Under-Specified Requirements} Certain vague or under-specified requirements within RPKI RFCs can lead to inconsistent implementations and operational uncertainty.  
The lack of clear guidance in areas like certificate handling, ROA validation, error handling, and manifest management creates potential windows of vulnerability, where different networks might adopt conflicting behaviors. This can increase the risk of security issues such as prefix hijacks or operational disruptions.

{\bf Discrepancies in Processing Certificates.} One example for vague specification is the handling of optional unknown field types within certificates. While the RPKI certificate template mandates the existence (or non-existence) of certain fields, e.g., mandating that the {\tt signerIdentifer} is present, it leaves open how validation should handle unknown optional fields. The X.509 template, however, allows for a range of additional optional fields not specified in the RPKI standard, and systems may decide to use such fields for specific use-cases. This problem already manifests in real-world RPKI operation. In previous research \cite{mirdita2023cure}, we identified that one of the four major relying parties, Fort, rejects objects with unknown optional fields in certificates, while all others ignore the field and still validate the object. We found that Amazon used an optional name field within their certificates, not specified within the RPKI standard. While all other relying parties validated the object, Fort rejected all such Amazon objects, leading to over 6000 prefixes by Amazon vulnerable to hijack in any system using Fort\footnote{\url{https://labs.ripe.net/author/niklas-vogel/crashing-the-party-vulnerabilities-in-rpki-relying-party-software/}}. After raising the issue to developers, there was disagreement between different implementations how to interpret the lack of specification in the standard. 

{\bf Discrepancies in Key Rollover.} A further example is guidelines for key rollover in RPKI in RFC6489, which leaves operational details open to interpretation. Specifically, it does not fully address how frequently key rollovers should occur or how long old keys should remain valid. In practice, this leads to implementations using different key rollover timelines, which can result in operational inconsistencies. A lack of detailed guidance on timing creates the risk that some network operators may perform key rollovers too quickly, leading to temporary route validation issues, while others may retain old keys for too long, increasing the vulnerability window for compromised keys. Another example is RFC6486, which covers manifests (used to ensure repository consistency), but does not clearly define how often manifests should be updated or what happens if a manifest becomes temporarily unavailable. It also lacks precise guidance on how to handle missing objects in the manifest. If a manifest is unavailable for a short time or if certain objects listed in the manifest are missing from the repository, RFC6486 leaves it unclear whether operators should reject all repository data, trust older manifest versions, or temporarily proceed without validation. This ambiguity could lead to inconsistent behaviors across networks, with some rejecting valid routes due to temporary manifest issues, while others continue to operate without validation.

{\bf Discrepancies in Thresholds.} Another example is the requirements to impose thresholds on the functionalities of the relying parties. While standards need to offer some flexibility of implementation to allow for competition among products meeting the standard, it is also important to avoid standards that wind up allowing implementations that defeat the purpose of the standard. The issue of thresholds and limits on computation of the relying parties was not clearly specified. This vagueness was shown to expose the relying parties to stalling attacks by malicious repositories \cite{DBLP:conf/uss/HlavacekJMSW22,DBLP:conf/sigcomm/HlavacekJMSW23}.
In the course of the attack, an adversary can hang RPKI validating resolvers for significant amounts of time, thereby stalling validation long enough for RPKI protection to be downgraded in BGP routers. While these problems have been addressed by adding thresholds, research shows that thresholds in relying parties \cite{DBLP:conf/sigcomm/HlavacekJMSW23} not only do not eliminate the attacks against RPKI validation, but worse, they introduce failures under benign network conditions. Specifically, on the one hand, the current thresholds are too permissive to prevent attacks and much stricter thresholds are required to enhance resilience against attacks. On the other hand, even the permissive thresholds may cause the relying parties to fail when fetching RPKI objects. A measurement we carried out in 2023 indicates that about 11.78\% of the attempts to fetch RPKI objects fail due to thresholds kicking in and dropping the connection \cite{DBLP:conf/sigcomm/HlavacekJMSW23}. 

Source code analysis reveals that the implemented threshold values differ significantly across the relying party implementations. For example, the connection timeout parameter for different implementations can range from 60s to 2.9h. These large discrepancies indicate that the developers had to rely on their intuition due to lack of official guidance for the selection of the thresholds values. Indeed, we have not found an analysis of the effect of different threshold values nor a guidance on their selection. The lack of specific requirements in the standards for thresholds and lack of analysis on the ramifications of thresholds on RPKI validation under diverse network conditions cause the developers to select arbitrary values in their implementations. 

\section{Maturity of RPKI Software Packages}\label{sc:software}

Since the introduction of RPKI, implementations of RPKI software have emerged as open-source projects, developed and managed by a small group of developers. Looking at the contribution history of the respective projects, the amount of active contributors ranges from 1 to 5 people\footnote{We define active contribution as contributing code in the last 12 months.}. All projects are open-source and accept pull requests, with one implementation specifically encouraging pull requests to the repository. 
While open-source software is beneficial to improve security, public participation, and trustworthiness of software, the XZ-Utils attack \cite{xzutils} has illustrated that especially projects managed by a small team of developers may be subject to intentional introduction of malicious code from a sophisticated attacker. Three implementations are developed by known organizations, indicating more available resources for security management and vetting of code contributions to the project, with organizations located in North America, South America, and Europe.

\subsection{Software Bugs} While the quality of implementations has continuously improved over recent years, we find that implementations repeatedly exhibit large amounts of vulnerabilities, with some problems persisting to-date.
The RPKI specification, the RPKI software packages and RPKI repository implementations are still not sufficiently stable, contain critical vulnerabilities and RPKI developers lack tools to test their software to identify bugs and vulnerabilities \cite{DBLP:conf/uss/HlavacekJMSW22,mirdita2022poster,DBLP:conf/sigcomm/HlavacekJMSW23,mirdita2023cure,cattepoel2024poster,jacobsen2024poster}.
Although vulnerabilities were discovered in all RPKI components, we find that the largest amount of vulnerabilities have been presented in implementations of relying party validators, the middleware responsible for downloading, parsing, and validating RPKI objects. Overall, at least 53 vulnerabilities were disclosed, including persistent DoS, authentication bypass, cache poisoning, and remote-code-execution. While the large majority of these vulnerabilities were swiftly fixed, they still raise the question on the resilience of implementations and potential existence of other zero days.

Previous work has indicated multiple reasons for the persistence of vulnerabilities and problems in RPKI implementations, naming a lack of test-tooling, complexity of the cryptographic architecture, and vagueness of RFC requirements. Further contributing to the problem, two of the three largest implementations of RPKI relying parties are written in C, a non memory-save language that is more difficult to protect against memory-related exploitation. 
While we expect that software security will increase in the future, with improved secure coding patterns and increasing availability of tooling, the current state of software security in RPKI makes it attractive for attackers, with relative abundance of vulnerabilities that have potentially devastating consequences for RPKI validation and might even open a back-door into the local network running the vulnerable software component.

\subsection{Intentional Backdoors} As RPKI is gaining traction, the risk of intentional backdoors will also grow. Following the exposure of the XZ-Utils attack \cite{xzutils}, the issue of planted backdoors in open-source software has gained elevated attention by the security community. Since all popular RPKI software implementations are open-source and accept code contributions by the community, the threat of intentional backdoors is substantial in the context of RPKI. Such backdoors might, for example, include compromises to validation integrity to get malicious data validated, eavesdropping on local networks, or even manipulated program flow to enable remote code execution (RCE). Contributions to open-source projects should thus be vetted by the developers, and the community should invest resources to track changes to the repositories and investigate suspicious changes to the software. 

\section{Maturity of RPKI Deployments}\label{sc:deployments}
The challenges in deploying RPKI evolve around errors with registering network resources in RPKI and lack of experience with strict RPKI validation in production environments. We explain these issues next.

\subsection{Errors in ROAs}
The deployment of RPKI involves creating ROAs for authenticating the associated network resources. The objects should be stored in public RPKI repositories. 
Managing ROAs and maintaining cryptographic keys introduces complexity. Research showed that complexity often leads to errors and misconfigurations in RPKI objects, which result in route filtering and loss of connectivity for legitimate routes \cite{DBLP:conf/ndss/GiladCHSS17}. Despite awareness to the issue of erroneous ROAs, and the fact that ROV enforcing networks filter traffic from prefixes which have conflicts with ROAs, according to the NIST RPKI monitor\footnote{\url{https://rpki-monitor.antd.nist.gov/}} erroneous ROAs still exist. Although there are proposals to automate registration of prefixes \cite{DBLP:conf/ndss/HlavacekCGHKSS20} no mechanisms are used in practice.

\subsection{Test-Mode ROV Deployments}
Networks set up and use relying parties to fetch and validate the ROAs and other RPKI material. The validated objects inform the routers in their routing decisions, in a process called ROV. 

{\bf Fail-Open Mode.}
Enforcement of RPKI validity with ROV is currently performed in test mode. RFC7115 recommends that operators' policies should not be too strict: the operators should use RPKI to prefer valid announcements, assign a lower preference to NotFound announcements, and either discard Invalid announcements or give them a very low preference.
To support this test mode, RPKI was designed to be ``fail open'', namely, if ROAs cannot be fetched, e.g., because they do not exist, or the RPKI repository that hosts them is unreachable, the RPKI validation for those resources gets the status ``Not Found'' and announcements are hence accepted. Operating RPKI validation in test mode is critical for stability of the Internet, since not all address space is covered by ROAs yet, and adversaries may be able to prevent relying party validators from fetching RPKI objects \cite{DBLP:conf/uss/HlavacekJMSW22}. In fact, failures to access the RPKI repositories may occur even under benign network conditions \cite{DBLP:conf/sigcomm/HlavacekJMSW23}. If strict validation is applied, prefixes, for which RPKI objects could not be retrieved will be filtered, hence impairing reachability to those ASes. The risk of losing legitimate traffic is a substantial concern for network operators and one of the main obstacles hindering wide adoption of RPKI validation. 
To mitigate the concerns for potential loss of legitimate traffic RPKI was designed to support operation in ``fail open`` mode. Operating in fail-open test mode facilitates incremental RPKI deployment, reducing failures and traffic loss. The downside of the fail open test mode is that networks that invest in deploying RPKI filtering with ROV may still be vulnerable to routing hijacks if adversaries can disable RPKI validation, e.g., by blocking access to the RPKI repositories or by preventing the relying parties from fetching fresh RPKI objects \cite{DBLP:conf/uss/HlavacekJMSW22}. As a result, adversaries may be able to hijack network resources covered with ROAs, if the ROV filtering is enforced in fail open mode, since the routers do not drop such invalid routes. 

Surely the fail-open mode does not offer sufficient security guarantees for Internet routing and in the long term there should be a transition to strict RPKI validation. However, enforcing strict validation, which RFC7715 acknowledges to not be realistic in the near future, exposes the networks to DoS attacks. For example, if existence of a valid ROA is mandated but a router does not have the required RPKI data, it will eventually drop all announcements, leading to full DoS and a lack of reachability to those routes. Thus, while the impact of attacks on availability differs between fail-open and strict validations, availability of RPKI data remains a core concern in RPKI deployments that still needs to be addressed. In addition, there should be a plan for transition to operating RPKI in production environments in strict validation mode, to identify any possible challenges and problems. Only after the community collects sufficient experience operating full fledged RPKI in strict validation mode can RPKI be established as a mature technology.

{\bf Experimental Setup.}
In addition to operation in a fail-open mode, recommendations generally suggest to start with an experimental setup. The FCC report recommends that networks deploy RPKI filtering in a staged process, starting with an experimental setups that do not impact their routing. MANRS also recommends that networks first install RPKI with logging enabled without actually filtering routes. As a result, many adopters do not prioritize RPKI, remaining in an experimental stage and do not gain experience with full-fledged production RPKI management. 

The existing recommendations to start by deploying RPKI without impacting real-world routing illustrates that the community needs more experience with deploying and operating RPKI validation. By recommending staged and experimental deployment, many systems do not move RPKI to the stage where they apply filtering and treat RPKI as a full production deployment, e.g., illustrated by a lacking patch-management of the RPs, and many systems issuing ROAs but not enforcing ROV in their routers. It is vital that more experience in RPKI operation, including impact of strict validation, is published, and more research into operational considerations of RPKI is conducted to provide a more solid understanding of RPKI deployments. Such research will guide the standardization, developers and operators towards a mature and ready-to-implement RPKI that can be deployed faster and more securely.

\subsection{Challenges in Full Deployment}
Despite the continuous growth of deployment, the scenario of a full global RPKI deployment is not yet well understood, and simulations for scalability of RPKI in different deployment scenarios were not sufficiently explored. This is important since research shows multiple indications for potential problems in a full deployment setting \cite{DBLP:conf/sigcomm/HlavacekJMSW23}. The growth in the number of RPKI objects and standardization of additional objects will significantly increase download and validation times for relying parties, and will eventually increase update intervals, extending the delay between changes in objects and updates to the routing behavior of systems. This will also decrease the agility of the systems to quickly react to problems and attacks in routing. Further, with increasing deployment also in smaller systems, the number of relying parties will increase, adding substantial additional load to the RPKI repositories. 

With more operational experience in RPKI, and under pressure of more attacks on the architecture, more systems might also decide to operate their own RPKI repository, significantly increasing the amount of RPKI repositories. Since more repositories require additional sessions from the RPs, and downloading of more data, this will additionally increase download times for RPs, further increasing the minimal possible update interval. All these challenges need to be understood and resolved to make a full global deployment of RPKI feasible.

\subsection{Challenges with Patch Management}
Many operators struggle with keeping RPKI software updated due to the lack of automated patching mechanisms. Getting the most up-to-date relying party software requires the system administrator to manually get and install the new version. We find that a significant percentage of RPKI installations run outdated versions with known vulnerabilities, increasing the risk of attacks. To measure how many relying parties were not patched, we set up our RPKI publication point and log version headers of global relying parties that contact our repository to fetch RPKI objects. We check how many clients use software with at least one publicly known severe vulnerability (CVSS >= 7.0). We only include vulnerabilities disclosed more than 1 year before our measurements. We find 41.2\% of global relying parties are vulnerable to at least one long-disclosed attack. This number is alarming as it shows that even when vulnerabilities are discovered and patched, many relying parties can still be attacked long after the patch was issued. The lack of timely patching also illustrates that management of relying parties is not prioritized. % by administrators.

\subsection{RPKI Operation}\label{sc.inconsistencies}
Automation tools for RPKI configuration and management are still developing. Many operators rely on manual processes, which can lead to misconfigurations. 

But, operational challenges arise also after the RPKI components have been configured and set up. A recent study \cite{dona:usenix2025} discovered a large and persistent amount of networking errors in RPKI repositories, including unreachability, slow connections, and failing DNSSEC validation, frequently leading to unavailable objects and slowing down the fetching of RPKI objects for all globally running relying parties. 
Another study showed that inconsistency of RPKI validation also poses challenges to RPKI deployments \cite{mirdita2023cure}. With RPKI utilizing established cryptographic algorithms, it would be expected that different RPKI implementations reach consistent validation results on identical objects. This is not the case in practice. 

In a follow-up research we carried out, we identified 25 inconsistencies that lead to discrepancies in RPKI validation results across different relying parties. Multiple of the discovered inconsistencies lead to differing validation results of real-world RPKI objects, therefore impacting production Internet routing \cite{ripearticle, fortcve,fortrelease,clientrelease,mirdita2023cure}. For instance, some of the implementations do not accept repositories that contain unknown object types, like ASPAs, or do not accept Snapshots where the same object was added twice. Some implementations do not accept objects with identical SKIs, which can occur in the case of improper key management. 
This inconsistency can cause legitimate routes to be incorrectly marked as invalid, leading to instability in routing decisions across different networks.

%% file: attacks.tex
\section{Emerging RPKI Attack Surface}\label{sc:surface}

The expanding adoption of RPKI in global networks also increases its attractiveness as a target for attacks. Therefore, it is important to extend the White House Roadmap to also consider attacks against RPKI, and attacks through RPKI against the RPKI-adopters. Such RPKI attacks may be closely directed towards RPKI functionality, e.g., downgrading or circumventing hijack protections in victim systems, but can also target RPKI components as entry-points into the network where vulnerable RPKI software is located. Enumeration of potential targets and development of directed attacks is straightforward, as IP addresses of relying parties and RPKI repositories are known, and they continuously communicate with others hosts in the Internet. Using enumerated targets, adversaries can attack the relying party instances directly by setting up their own RPKI repository, or indirectly by attacking other RPKI repositories. Attacks often use specially crafted RPKI objects that disable or manipulate RPKI validation or, in the worst case, allow the attacker to obtain access to the RPKI-adopting network. In this section, we discuss the RPKI-specific and general attack surface of networks.

\begin{table}[t!]
\centering
\renewcommand{\arraystretch}{0.9}
\scriptsize
\begin{tabular}{|l|c|c|c|c|c|c|}
\hline
\textbf{Vulnerability}       & \textbf{Amount} & \textbf{Target} & \textbf{Availability} & \textbf{RCE} & \textbf{Validation } & \textbf{CVSS} \\ %\hline
\textbf{}       & \textbf{} & \textbf{} & \textbf{} & \textbf{} & \textbf{ Integrity} & \textbf{} \\ \hline\hline
Crash                        & 42             & RP              & \checkmark   & --           & --                        & 7.5  \\ \hline
Stalling                    & 3              & PP              & \checkmark   & --           & --                        & N.A. \\ \hline
Kill-switch                  & 1              & RRDP            & \checkmark   & --           & --                        & 7.5  \\ \hline
PDU                     & 1              & RTR             & \checkmark   & --           & --                        & N.A. \\ \hline
Info-Leak                     & 1              & PP             & --   & --           & --                        & N.A. \\ \hline
Delta-Snapshot                     & 1              & PP             & --   & --           & --                        & N.A. \\ \hline
Cache-Poisoning              & 1              & RP              & \checkmark   & --           & \checkmark                & N.A. \\ \hline
Path-Traversal               & 2              & RP              & \checkmark   & \checkmark   & \checkmark                & 9.3  \\ \hline
Buffer-Overflow              & 1              & RP              & \checkmark   & \checkmark   & \checkmark                & 9.8  \\ \hline
\end{tabular}
\vspace{-1pt}
\caption{Published vulnerabilities in RPKI components.}
\vspace{-15pt}
\label{tab:vulnerabilities}
\end{table}

\subsection{Downgrade of RPKI Validation}
RPKI validation of BGP messages requires availability of RPKI data. We find that attacks on availability are the most prominent threat in the RPKI ecosystem, with at least 53 such vulnerabilities published in scientific literature within the last 3 years \cite{DBLP:conf/uss/HlavacekJMSW22,DBLP:conf/ccs/HlavacekJMSW22,DBLP:conf/sigcomm/HlavacekJMSW23, van2022rpkiller, mirdita2023cure, jacobsen2024poster,mirdita2022poster}. Attacks on availability may target all parts of the RPKI architecture, with attacks ranging from rate-limiting RPKI repositories or creating complex repository structures to stall the relying party validators, to crashing the validators through malformed objects or protocol headers. We summarize the known vulnerabilities against RPKI components in Table \ref{tab:vulnerabilities}.

{\bf Attacks Against Relying Parties.} Malicious RPKI repositories can exploit vulnerabilities in the RPKI protocol specification or in software implementations to stall relying parties or crash them, preventing them from downloading and validating ROAs. As listed in Table \ref{tab:vulnerabilities}, 42 vulnerabilities have already been disclosed that could be exploited to crash relying parties during processing, one crashing through RRDP and 3 vulnerabilities that compromised availability through stalling. The main goal of attacks against availability is to prevent RPKI validation. 

{\bf Attacks Against Repositories.} 
To prevent relying parties from accessing RPKI data, adversaries can launch DoS attacks against repositories, exploiting vulnerabilities in repositories software \cite{cattepoel2024poster} and causing them to crash, or flooding them with requests \cite{DBLP:conf/ccs/HlavacekJMSW22}.

{\bf Effect.} 
If a border router cannot fetch fresh RPKI data from a relying party, it will not use RPKI for making routing decisions in BGP, i.e., its RPKI protection will be downgraded to insecure BGP. 
Default routers cache flush times range from 360s (Cisco \cite{cisco:rpki}) to 7200s (FRR \cite{frr:rpki}) if a relying party is unavailable. Further, if a relying party cannot fetch objects due to stalling, cached objects will eventually expire, generally within a maximum of 24h. While handling of stale objects depends on local policy, most implementations will eventually discard stale objects, and protection is thus downgraded. As a result, when the networks cannot apply RPKI validation, they are exposed to routing attacks on supposedly RPKI-protected systems. With increasing deployment of RPKI, including networks that have been historically attractive for cyberattacks like DNS servers or ASes hosting crypto-currency services, attacks on availability become more attractive.

\subsection{Manipulating RPKI Validation Integrity}
While attacks on availability of the relying parties or the repositories downgrade RPKI protection, attacks against integrity aim to manipulate RPKI validation results to circumvent protection. Multiple vulnerabilities have been disclosed in recent years that allow an attacker to subvert validation integrity.

{\bf Attacks Against Relying Parties.} Existing attacks allow to either get malicious objects validated, or get specific RPKI objects invalidated, see Table \ref{tab:vulnerabilities}. The former attack can be done by adding a malicious Trust Anchor Locator (TAL) via a path-traversal attack against relying parties and allows an attacker to introduce malicious RPKI objects \cite{mirdita2023cure}. The attacker can therefore attest ownership of arbitrary IP resources and thus conduct BGP attacks that seem like RPKI-valid announcements. The latter attack targets specific objects within the RPKI, and has also been first demonstrated by \cite{mirdita2023cure}. For example, a vulnerability within one relying party validator allowed an attacker to issue an object with a specific key identifier to invalidate any other object within the RPKI that has the same identifier. This kind of targeted attack allows the attacker to downgrade protection only for specific resources, while not impacting availability or integrity of any other resource.

{\bf Attacks Against RPKI Repositories.} 
Repositories can be attacked, e.g., over a management interface, which is exposed to the Internet, allowing attackers to gain access to the management tooling when weak or default credentials are used \cite{DBLP:conf/uss/DaiJSW21}. 

{\bf Effect.} These attacks are very difficult to detect and are thus much more attractive for attackers than attacks on availability, though zero-days are likely much harder to find, with only two such vulnerabilities published to-date. 

\subsection{RPKI As a Foothold Into The Network}
The attack surface of networks deploying RPKI is not limited to attacks on RPKI functionality. RPKI components, particularly relying party instances, can expose networks to a wider range of attacks. 
It is recommended to set up a relying party in proximity to the border routers, ideally on the same network, see Best Current Practices (BCP) BCP-185 (also known as RFC7115). Consequently, any intentional backdoors or erroneous bugs in a relying party software can be exploited by adversaries to attack the border routers and even penetrate the network. 

{\bf RCE in Fort.}
We explain this risk on an example of a Remote Code Execution (RCE) attack against RPKI, which we discovered in our research \cite{jacobsen2024poster}. Concretely in this example we discovered a buffer-overflow vulnerability in the validation pipeline of Fort relying party validator. The vulnerability stems from a bug in the processing of the key-usage extension of an X.509 certificate, contained in most RPKI objects. The 9 bit field differentiates the usage of keys within the certificates, indicating if the key can be used to sign other certificates, or RPKI payload objects. Since the extension usually does not exceed a length of 2 bytes, the implementation allocates a 2 byte buffer to store the extension value. After allocating the buffer, the value is copied into the 2 byte buffer on the stack, without checking the actual length of the data. If an attacker inserts more than 2 bytes, and ensures the first 9 bit conform to the requirements for the key-usage extension, he can overflow the stack and write arbitrary bytes into stack memory behind the allocated buffer. The attack can be exploited by setting up a production RPKI repository, uploading an object with a manipulated key-usage extension, and serving it to any client accessing the repository content. Through the vulnerability, the attacker can overwrite the stack, manipulate the control-flow of Fort and thereby achieve RCE on all clients running the software. 
The severity of the bug raises questions on its origin, and we find potential indications for both intentional introduction of the bug and a benign coding error.
First, the code section does not implement any functionality, putting into question why the code was added in the first place. Further, the convenient accessibility of the vulnerability from any remote RPKI repository makes exploitation easy and stealthy. Also, the vulnerability can be activated in new distributions of the software by simply changing some compilation flags, which will likely not be noticed by users. These observations might indicate malicious intent. However, we do not find any operational indications for malicious planting of the vulnerability. The code was committed by the long-term main developer of the software, and similar code-sections exist in other parts of the relying party that implement actual functionality, making a copy-paste error likely. While there is no clear answer on the intention, the ease with which such a backdoor could be inserted should raise awareness to the increase in the attack surface and in particular the risk of backdoors that RPKI introduces.

{\bf Effect.}
RCE attacks \cite{mirdita2023cure,jacobsen2024poster,van2022rpkiller} have very severe impact as they open the victim up for a wide range of follow-up exploitation. An attacker can, e.g., establish a reverse shell to the victim relying party or the repository, gain initial access, and infiltrate both the system itself and use it as a backdoor into the local network to attack other servers, like BGP routers. Further, the attacker might find sensitive information stored on the relying party or the repository, which might, e.g., include router credentials. Further, gaining access to the RPKI component also compromises validation integrity, as the attacker can add arbitrary data to the unauthenticated RPKI payload list that is sent to the routers without any additional validation.

%% file: conclusion.tex
\section{Perfect is the Enemy of the Good}\label{sc:role}

Until recently, few people outside the Internet operational, engineering, and research communities were aware of RPKI. That changed in September 2024 when the White House identified RPKI as the key component for securing Internet routing, pushing RPKI from niche to mainstream. It may be expected that mainstream technologies are fully mature, in particular, stable and secure. As a niche technology, RPKI developed organically in many small steps, each inching a bit closer to maturity. But, as the  previous sections showed, RPKI is far from being fully mature. What does this actually mean in practice? Did the White House push for the adoption of an immature technology, potentially doing more harm than good? Or did the White House promote the best available, good enough technology, motivating research and industry to speed up and put more resources behind improving RPKI?

{\bf Academic analysis shows RPKI is not mature.}
The academic exercise of applying the Technology Readiness Levels \cite{mankins1995technology,tyson2003interpreting} to RPKI indicates that RPKI is still in the stage of demonstration of system prototype in a test mode in an operational environment, i.e., below TRL 9. The RPKI implementations are not sufficiently stable and lack resilience to existing and future cyberattacks. The RPKI validation exhibits inconsistent results. The RPKI standard specifications have not yet been finalized. The developers and operators lack documentation and automated tools for development and configuration of RPKI technology. All these indicate that RPKI is not  mature.

{\bf But, so what? Systems in the real world are never fully mature.} Arguably, demanding full maturity before large-scale deployment is a very academic expectation; in real life, there is nothing like full maturity and perfection, only more or less good enough. The Internet, like most information and communication technologies, is not mature, from applications to Internet protocols and to popular security mechanisms, e.g., IPsec and SSL/TLS. All are vulnerable and contain bugs. Many Internet systems started from collaborative efforts between researchers and operators and grew organically. Over time these efforts mature from experimental research prototypes and individual initiatives into deployments by large networks. The software is improved `on-the-fly' with periodic patches, that close bugs or add new features. This organic systems maturity is not aligned with the academic definitions and frameworks, but in real life, the systems are never 100\% perfect and mature, but rather evolve gradually.

{\bf Mature or not, BGP connects the Internet.} Examples of immature, but nevertheless heavily used technologies exist in abundance; Internet routing with BGP is among the most prominent ones. Nowadays BGP enables all Internet activities.
In addition to its central role to any online activity, the complexity of BGP also grew. BGP was designed on three napkins, to connect different Internet domains. BGP was not designed to be a robust and security protocol that one would rely on for critical functionalities like the Internet has become. BGP has since then evolved, in the number of steps of its decision process, the number of attributes, the number of networks it supports, all these got more complicated, difficult to configure, vulnerable. Indeed, software bugs and issues in protocol specification are common in inter-domain routing with BGP and may lead to outages, failures and attacks. For instance, FRR routers crashed and disconnected large networks from the Internet because they could not parse standard-compliant BGP attributes in routing announcements\footnote{\url{https://www.zdnet.com/article/internet-experiment-goes-wrong-takes-down-a-bunch-of-linux-routers/}}. In addition, the complexity of BGP may create a chain of side effects, such that a small failure or misconfiguration in one part of the Internet can have devastating global consequences.

Despite all the problems, outages and attacks, the triple napkin protocol connects the Internet. Not only that, but also the applications of BGP evolved far beyond BGP's original purpose, including many new and emerging applications, such as internal routing within data centers, MPLS VPN across organizational sites, load balancing, and more. 

 {\bf Our analysis should be used as a TODO list.} An academic analysis is important and it allows to identify directions to improve security and stability of systems, but {\em the implementation of academic analysis needs to be adapted to how the systems evolve and mature in the real world}. An academic analysis provides a {\em TODO} list to guide the adopters, operators and developers in prioritizing their actions, addressing the problems one at a time, towards improving the maturity of an operating system. The list of problems however does not reflect the state of maturity of a system. 

{\bf The roadmap is a huge leap forward.}
The White House's  2024 roadmap's recognition of RPKI as a critical security measure is an important step forward. Until recently RPKI was mostly experimental, but the cyber security strategy of the white-house, the NPRM of the FCC, and the recent roadmap, make a huge leap forward, towards securing the routing infrastructure with RPKI. This roadmap will push the adoption of RPKI forward. Now it is important to identify the hurdles that need to be resolved towards this goal. We outlined a number of such challenges that need to be addressed in this work.

\section{Conclusions}\label{sc:conclusions}

RPKI implementations started as collaborative efforts between researchers, operators, and the broader IETF community. Over time, these efforts matured from experimental research projects and individual operator initiatives into deployments by some of the largest networks in the world. Our research shows that RPKI still suffers from problems and is not sufficiently stable. Nevertheless, RPKI already delivers benefit and it is an essential part of the Internet's ongoing efforts to improve routing security. Research shows that RPKI can substantially limit the propagation of invalid BGP announcements, hence mitigating traffic hijacks \cite{hlavacek2023keep}. RPKI also provides an important prerequisite for prospective routing security solutions, including origin validation, path-validation, and route leak prevention. The roadmap of the White-House is a huge leap for RPKI, and therefore also for Internet routing, to truly mature and meet the expectations of security, reliability, and scalability for production-level deployments across the global Internet.

%% file: main.bbl
%%% -*-BibTeX-*-
%%% Do NOT edit. File created by BibTeX with style
%%% ACM-Reference-Format-Journals [18-Jan-2012].

\begin{thebibliography}{00}

%%% ====================================================================
%%% NOTE TO THE USER: you can override these defaults by providing
%%% customized versions of any of these macros before the \bibliography
%%% command.  Each of them MUST provide its own final punctuation,
%%% except for \shownote{}, \showDOI{}, and \showURL{}.  The latter two
%%% do not use final punctuation, in order to avoid confusing it with
%%% the Web address.
%%%
%%% To suppress output of a particular field, define its macro to expand
%%% to an empty string, or better, \unskip, like this:
%%%
%%% \newcommand{\showDOI}[1]{\unskip}   % LaTeX syntax
%%%
%%% \def \showDOI #1{\unskip}           % plain TeX syntax
%%%
%%% ====================================================================

\ifx \showCODEN    \undefined \def \showCODEN     #1{\unskip}     \fi
\ifx \showDOI      \undefined \def \showDOI       #1{#1}\fi
\ifx \showISBNx    \undefined \def \showISBNx     #1{\unskip}     \fi
\ifx \showISBNxiii \undefined \def \showISBNxiii  #1{\unskip}     \fi
\ifx \showISSN     \undefined \def \showISSN      #1{\unskip}     \fi
\ifx \showLCCN     \undefined \def \showLCCN      #1{\unskip}     \fi
\ifx \shownote     \undefined \def \shownote      #1{#1}          \fi
\ifx \showarticletitle \undefined \def \showarticletitle #1{#1}   \fi
\ifx \showURL      \undefined \def \showURL       {\relax}        \fi
% The following commands are used for tagged output and should be
% invisible to TeX
\providecommand\bibfield[2]{#2}
\providecommand\bibinfo[2]{#2}
\providecommand\natexlab[1]{#1}
\providecommand\showeprint[2][]{arXiv:#2}

\bibitem[\protect\citeauthoryear{Austein, Bellovin, and Bush}{Austein et~al\mbox{.}}{2010}]%
        {nanog:history2}
\bibfield{author}{\bibinfo{person}{Rob Austein}, \bibinfo{person}{Steve Bellovin}, {and} \bibinfo{person}{Randy Bush}.} \bibinfo{year}{2010}\natexlab{}.
\newblock \bibinfo{title}{{The RPKI and Origin Validation}}.
\newblock \bibinfo{howpublished}{\\\url{https://conference.apnic.net/29/pdf/Routing-Security_02_The-RPKI-Origin-Validation_Randy-Bush.pdf}}.   (\bibinfo{year}{2010}).
\newblock
\newblock
\shownote{NANOG Archive.}


\bibitem[\protect\citeauthoryear{Azimov, Bogomazov, and Bush}{Azimov et~al\mbox{.}}{2024}]%
        {asparfc}
\bibfield{author}{\bibinfo{person}{A Azimov}, \bibinfo{person}{E Bogomazov}, {and} \bibinfo{person}{R Bush}.} \bibinfo{year}{2024}\natexlab{}.
\newblock \bibinfo{title}{BGP AS\_PATH Verification Based on Autonomous System Provider Authorization (ASPA) Objects}.
\newblock   (\bibinfo{year}{2024}).
\newblock


\bibitem[\protect\citeauthoryear{Bush}{Bush}{2014}]%
        {bush2014rfc}
\bibfield{author}{\bibinfo{person}{R Bush}.} \bibinfo{year}{2014}\natexlab{}.
\newblock \bibinfo{title}{RFC 7115: Origin Validation Operation Based on the Resource Public Key Infrastructure (RPKI)}.
\newblock   (\bibinfo{year}{2014}).
\newblock


\bibitem[\protect\citeauthoryear{Bush and Austein}{Bush and Austein}{2017}]%
        {bush2017resource}
\bibfield{author}{\bibinfo{person}{Randy Bush} {and} \bibinfo{person}{Rob Austein}.} \bibinfo{year}{2017}\natexlab{}.
\newblock \bibinfo{title}{The resource public key infrastructure (RPKI) to router protocol, version 1}.
\newblock   (\bibinfo{year}{2017}).
\newblock


\bibitem[\protect\citeauthoryear{Bush, Elkins, Austein, and Smith}{Bush et~al\mbox{.}}{2011}]%
        {nanog:history}
\bibfield{author}{\bibinfo{person}{Randy Bush}, \bibinfo{person}{Michael Elkins}, \bibinfo{person}{Rob Austein}, {and} \bibinfo{person}{Philip Smith}.} \bibinfo{year}{2011}\natexlab{}.
\newblock \bibinfo{title}{{RPKI Workshop Agenda}}.
\newblock \bibinfo{howpublished}{\\\url{https://archive.nanog.org/meetings/nanog52/presentations/Sunday/110612.nanog-lab-agenda.pdf}}.   (\bibinfo{year}{2011}).
\newblock
\newblock
\shownote{NANOG Archive.}


\bibitem[\protect\citeauthoryear{Butler, Farley, McDaniel, and Rexford}{Butler et~al\mbox{.}}{2009}]%
        {butler2009survey}
\bibfield{author}{\bibinfo{person}{Kevin Butler}, \bibinfo{person}{Toni~R Farley}, \bibinfo{person}{Patrick McDaniel}, {and} \bibinfo{person}{Jennifer Rexford}.} \bibinfo{year}{2009}\natexlab{}.
\newblock \showarticletitle{A survey of BGP security issues and solutions}.
\newblock \bibinfo{journal}{{\it Proc. IEEE}} \bibinfo{volume}{98}, \bibinfo{number}{1} (\bibinfo{year}{2009}), \bibinfo{pages}{100--122}.
\newblock


\bibitem[\protect\citeauthoryear{Cattepoel, Mirdita, Shulman, and Waidner}{Cattepoel et~al\mbox{.}}{2024}]%
        {cattepoel2024poster}
\bibfield{author}{\bibinfo{person}{Louis Cattepoel}, \bibinfo{person}{Donika Mirdita}, \bibinfo{person}{Haya Shulman}, {and} \bibinfo{person}{Michael Waidner}.} \bibinfo{year}{2024}\natexlab{}.
\newblock \showarticletitle{{Poster: Kill Krill or Proxy RPKI}}. In \bibinfo{booktitle}{{\em Proceedings of the 2024 ACM SIGSAC Conference on Computer and Communications Security}}.
\newblock


\bibitem[\protect\citeauthoryear{Cho, Fontugne, Cho, Dainotti, and Gill}{Cho et~al\mbox{.}}{2019}]%
        {cho2019bgp}
\bibfield{author}{\bibinfo{person}{Shinyoung Cho}, \bibinfo{person}{Romain Fontugne}, \bibinfo{person}{Kenjiro Cho}, \bibinfo{person}{Alberto Dainotti}, {and} \bibinfo{person}{Phillipa Gill}.} \bibinfo{year}{2019}\natexlab{}.
\newblock \showarticletitle{BGP hijacking classification}. In \bibinfo{booktitle}{{\em 2019 Network Traffic Measurement and Analysis Conference (TMA)}}. IEEE, \bibinfo{pages}{25--32}.
\newblock


\bibitem[\protect\citeauthoryear{Cimpanu}{Cimpanu}{2022}]%
        {klayswap}
\bibfield{author}{\bibinfo{person}{Catalin Cimpanu}.} \bibinfo{year}{2022}\natexlab{}.
\newblock \bibinfo{title}{{ KlaySwap crypto users lose funds after BGP hijack}}.
\newblock \bibinfo{howpublished}{\\\url{https://therecord.media/klayswap-crypto-users-lose-funds-after-bgp-hijack}}.   (\bibinfo{year}{2022}).
\newblock
\newblock
\shownote{Accessed 09/09/2024.}


\bibitem[\protect\citeauthoryear{Cisco}{Cisco}{2023}]%
        {cisco:rpki}
\bibfield{author}{\bibinfo{person}{Cisco}.} \bibinfo{year}{2023}\natexlab{}.
\newblock \bibinfo{title}{{ Routing Configuration Guide for Cisco ASR 9000 Series Routers, IOS XR Release 6.2.x }}.
\newblock \bibinfo{howpublished}{\\\url{https://www.cisco.com/c/en/us/td/docs/routers/asr9000/software/asr9k-r6-2/routing/configuration/guide/b-routing-cg-asr9000-62x/b-routing-cg-asr9000-62x_chapter_010.html}}.   (\bibinfo{year}{2023}).
\newblock


\bibitem[\protect\citeauthoryear{Dai, Jeitner, Schulmann, and Waidner}{Dai et~al\mbox{.}}{2021}]%
        {DBLP:conf/uss/DaiJSW21}
\bibfield{author}{\bibinfo{person}{Tianxiang Dai}, \bibinfo{person}{Philipp Jeitner}, \bibinfo{person}{Haya Schulmann}, {and} \bibinfo{person}{Michael Waidner}.} \bibinfo{year}{2021}\natexlab{}.
\newblock \showarticletitle{{The Hijackers Guide To The Galaxy: Off-Path Taking Over Internet Resources}}. In \bibinfo{booktitle}{{\em 30th {USENIX} Security Symposium, {USENIX} Security 2021, August 11-13, 2021}}, \bibfield{editor}{\bibinfo{person}{Michael~D. Bailey} {and} \bibinfo{person}{Rachel Greenstadt}} (Eds.). \bibinfo{publisher}{{USENIX} Association}, \bibinfo{pages}{3147--3164}.
\newblock


\bibitem[\protect\citeauthoryear{FRR}{FRR}{2023}]%
        {frr:rpki}
\bibfield{author}{\bibinfo{person}{FRR}.} \bibinfo{year}{2023}\natexlab{}.
\newblock \bibinfo{title}{{FRRouting}}.
\newblock \bibinfo{howpublished}{\\\url{https://docs.frrouting.org/en/latest/bgp.htmlconfiguring-rpki-rtr-cache-servers}}.   (\bibinfo{year}{2023}).
\newblock


\bibitem[\protect\citeauthoryear{Gao and Rexford}{Gao and Rexford}{2000}]%
        {gao2000stable}
\bibfield{author}{\bibinfo{person}{Lixin Gao} {and} \bibinfo{person}{Jennifer Rexford}.} \bibinfo{year}{2000}\natexlab{}.
\newblock \showarticletitle{Stable internet routing without global coordination}. In \bibinfo{booktitle}{{\em Proceedings of the 2000 ACM SIGMETRICS international conference on Measurement and modeling of computer systems}}. \bibinfo{pages}{307--317}.
\newblock


\bibitem[\protect\citeauthoryear{Gilad, Cohen, Herzberg, Schapira, and Schulmann}{Gilad et~al\mbox{.}}{2017}]%
        {DBLP:conf/ndss/GiladCHSS17}
\bibfield{author}{\bibinfo{person}{Yossi Gilad}, \bibinfo{person}{Avichai Cohen}, \bibinfo{person}{Amir Herzberg}, \bibinfo{person}{Michael Schapira}, {and} \bibinfo{person}{Haya Schulmann}.} \bibinfo{year}{2017}\natexlab{}.
\newblock \showarticletitle{Are We There Yet? On RPKI's Deployment and Security}. In \bibinfo{booktitle}{{\em 24th Annual Network and Distributed System Security Symposium, {NDSS} 2017, San Diego, California, USA, February 26 - March 1, 2017}}. \bibinfo{publisher}{The Internet Society}.
\newblock


\bibitem[\protect\citeauthoryear{Herdes, Zhang, and Ryan}{Herdes et~al\mbox{.}}{2024}]%
        {cloudfarehijack}
\bibfield{author}{\bibinfo{person}{Bryton Herdes}, \bibinfo{person}{Mingwei Zhang}, {and} \bibinfo{person}{Tanner Ryan}.} \bibinfo{year}{2024}\natexlab{}.
\newblock \bibinfo{title}{{ Cloudflare 1.1.1.1 incident on June 27, 2024}}.
\newblock \bibinfo{howpublished}{\\\url{https://blog.cloudflare.com/cloudflare-1111-incident-on-june-27-2024/}}.   (\bibinfo{year}{2024}).
\newblock
\newblock
\shownote{Accessed 09/09/2024.}


\bibitem[\protect\citeauthoryear{Hlavacek, Cunha, Gilad, Herzberg, Katz{-}Bassett, Schapira, and Schulmann}{Hlavacek et~al\mbox{.}}{2020}]%
        {DBLP:conf/ndss/HlavacekCGHKSS20}
\bibfield{author}{\bibinfo{person}{Tomas Hlavacek}, \bibinfo{person}{{\'{I}}talo Cunha}, \bibinfo{person}{Yossi Gilad}, \bibinfo{person}{Amir Herzberg}, \bibinfo{person}{Ethan Katz{-}Bassett}, \bibinfo{person}{Michael Schapira}, {and} \bibinfo{person}{Haya Schulmann}.} \bibinfo{year}{2020}\natexlab{}.
\newblock \showarticletitle{{DISCO:} Sidestepping RPKI's Deployment Barriers}. In \bibinfo{booktitle}{{\em 27th Annual Network and Distributed System Security Symposium, {NDSS} 2020, San Diego, California, USA, February 23-26, 2020}}. \bibinfo{publisher}{The Internet Society}.
\newblock


\bibitem[\protect\citeauthoryear{Hlavacek, Jeitner, Mirdita, Schulmann, and Waidner}{Hlavacek et~al\mbox{.}}{2022a}]%
        {DBLP:conf/ccs/HlavacekJMSW22}
\bibfield{author}{\bibinfo{person}{Tomas Hlavacek}, \bibinfo{person}{Philipp Jeitner}, \bibinfo{person}{Donika Mirdita}, \bibinfo{person}{Haya Schulmann}, {and} \bibinfo{person}{Michael Waidner}.} \bibinfo{year}{2022}\natexlab{a}.
\newblock \showarticletitle{Behind the Scenes of {RPKI}}. In \bibinfo{booktitle}{{\em Proceedings of the 2022 {ACM} {SIGSAC} Conference on Computer and Communications Security, {CCS} 2022, Los Angeles, CA, USA, November 7-11, 2022}}, \bibfield{editor}{\bibinfo{person}{Heng Yin}, \bibinfo{person}{Angelos Stavrou}, \bibinfo{person}{Cas Cremers}, {and} \bibinfo{person}{Elaine Shi}} (Eds.). \bibinfo{publisher}{{ACM}}, \bibinfo{pages}{1413--1426}.
\newblock


\bibitem[\protect\citeauthoryear{Hlavacek, Jeitner, Mirdita, Schulmann, and Waidner}{Hlavacek et~al\mbox{.}}{2022b}]%
        {DBLP:conf/uss/HlavacekJMSW22}
\bibfield{author}{\bibinfo{person}{Tomas Hlavacek}, \bibinfo{person}{Philipp Jeitner}, \bibinfo{person}{Donika Mirdita}, \bibinfo{person}{Haya Schulmann}, {and} \bibinfo{person}{Michael Waidner}.} \bibinfo{year}{2022}\natexlab{b}.
\newblock \showarticletitle{Stalloris: {RPKI} Downgrade Attack}. In \bibinfo{booktitle}{{\em 31st {USENIX} Security Symposium, {USENIX} Security 2022, Boston, MA, USA, August 10-12, 2022}}, \bibfield{editor}{\bibinfo{person}{Kevin R.~B. Butler} {and} \bibinfo{person}{Kurt Thomas}} (Eds.). \bibinfo{publisher}{{USENIX} Association}, \bibinfo{pages}{4455--4471}.
\newblock


\bibitem[\protect\citeauthoryear{Hlavacek, Jeitner, Mirdita, Schulmann, and Waidner}{Hlavacek et~al\mbox{.}}{2023a}]%
        {DBLP:conf/sigcomm/HlavacekJMSW23}
\bibfield{author}{\bibinfo{person}{Tomas Hlavacek}, \bibinfo{person}{Philipp Jeitner}, \bibinfo{person}{Donika Mirdita}, \bibinfo{person}{Haya Schulmann}, {and} \bibinfo{person}{Michael Waidner}.} \bibinfo{year}{2023}\natexlab{a}.
\newblock \showarticletitle{Beyond Limits: How to Disable Validators in Secure Networks}. In \bibinfo{booktitle}{{\em Proceedings of the {ACM} {SIGCOMM} 2023 Conference, {ACM} {SIGCOMM} 2023, New York, NY, USA, 10-14 September 2023}}, \bibfield{editor}{\bibinfo{person}{Henning Schulzrinne}, \bibinfo{person}{Vishal Misra}, \bibinfo{person}{Eddie Kohler}, {and} \bibinfo{person}{David~A. Maltz}} (Eds.). \bibinfo{publisher}{{ACM}}, \bibinfo{pages}{950--966}.
\newblock


\bibitem[\protect\citeauthoryear{Hlavacek, Shulman, Vogel, and Waidner}{Hlavacek et~al\mbox{.}}{2023b}]%
        {hlavacek2023keep}
\bibfield{author}{\bibinfo{person}{Tomas Hlavacek}, \bibinfo{person}{Haya Shulman}, \bibinfo{person}{Niklas Vogel}, {and} \bibinfo{person}{Michael Waidner}.} \bibinfo{year}{2023}\natexlab{b}.
\newblock \showarticletitle{Keep Your Friends Close, but Your Routeservers Closer: Insights into $\{$RPKI$\}$ Validation in the Internet}. In \bibinfo{booktitle}{{\em 32nd USENIX Security Symposium (USENIX Security 23)}}. \bibinfo{pages}{4841--4858}.
\newblock


\bibitem[\protect\citeauthoryear{Huston, Michaelson, and Loomans}{Huston et~al\mbox{.}}{2012}]%
        {x509rfc}
\bibfield{author}{\bibinfo{person}{Geoff Huston}, \bibinfo{person}{George Michaelson}, {and} \bibinfo{person}{Robert Loomans}.} \bibinfo{year}{2012}\natexlab{}.
\newblock \bibinfo{title}{RFC 6487: A profile for X. 509 PKIX resource certificates}.
\newblock   (\bibinfo{year}{2012}).
\newblock


\bibitem[\protect\citeauthoryear{(IETF)}{(IETF)}{2012}]%
        {rpkirfc}
\bibfield{author}{\bibinfo{person}{Internet Engineering~Taskforce (IETF)}.} \bibinfo{year}{2012}\natexlab{}.
\newblock \bibinfo{title}{{ An Infrastructure to Support Secure Internet Routing (RFC6480) }}.
\newblock \bibinfo{howpublished}{\\\url{https://datatracker.ietf.org/doc/html/rfc6480}}.   (\bibinfo{year}{2012}).
\newblock


\bibitem[\protect\citeauthoryear{Jacobsen, Schulmann, Vogel, and Waidner}{Jacobsen et~al\mbox{.}}{2024}]%
        {jacobsen2024poster}
\bibfield{author}{\bibinfo{person}{Oliver Jacobsen}, \bibinfo{person}{Haya Schulmann}, \bibinfo{person}{Niklas Vogel}, {and} \bibinfo{person}{Michael Waidner}.} \bibinfo{year}{2024}\natexlab{}.
\newblock \showarticletitle{{Poster: From Fort to Foe: The Threat of RCE in RPKI}}. In \bibinfo{booktitle}{{\em Proceedings of the 2024 ACM SIGSAC Conference on Computer and Communications Security}}.
\newblock


\bibitem[\protect\citeauthoryear{Lepinski and Kent}{Lepinski and Kent}{2012}]%
        {lepinski2012rfc}
\bibfield{author}{\bibinfo{person}{Matt Lepinski} {and} \bibinfo{person}{S Kent}.} \bibinfo{year}{2012}\natexlab{}.
\newblock \bibinfo{title}{RFC 6480: an infrastructure to support secure Internet routing}.
\newblock   (\bibinfo{year}{2012}).
\newblock


\bibitem[\protect\citeauthoryear{Lepinski, Kent, and Kong}{Lepinski et~al\mbox{.}}{2012}]%
        {roarfc}
\bibfield{author}{\bibinfo{person}{Matt Lepinski}, \bibinfo{person}{Stephen Kent}, {and} \bibinfo{person}{Derrick Kong}.} \bibinfo{year}{2012}\natexlab{}.
\newblock \bibinfo{title}{RFC 6482: A profile for route origin authorizations (ROAs)}.
\newblock   (\bibinfo{year}{2012}).
\newblock


\bibitem[\protect\citeauthoryear{Lepinski and Sriram}{Lepinski and Sriram}{2017}]%
        {lepinski2017bgpsec}
\bibfield{author}{\bibinfo{person}{Matt Lepinski} {and} \bibinfo{person}{Kotikalapudi Sriram}.} \bibinfo{year}{2017}\natexlab{}.
\newblock \bibinfo{title}{RFC 8205: BGPSEC protocol specification}.
\newblock   (\bibinfo{year}{2017}).
\newblock


\bibitem[\protect\citeauthoryear{Li, Lin, Ashiq, Aben, Fontugne, Phokeer, and Chung}{Li et~al\mbox{.}}{2023}]%
        {li2023rovista}
\bibfield{author}{\bibinfo{person}{Weitong Li}, \bibinfo{person}{Zhexiao Lin}, \bibinfo{person}{Md~Ishtiaq Ashiq}, \bibinfo{person}{Emile Aben}, \bibinfo{person}{Romain Fontugne}, \bibinfo{person}{Amreesh Phokeer}, {and} \bibinfo{person}{Taejoong Chung}.} \bibinfo{year}{2023}\natexlab{}.
\newblock \showarticletitle{RoVista: Measuring and analyzing the route origin validation (ROV) in RPKI}. In \bibinfo{booktitle}{{\em Proceedings of the 2023 ACM on Internet Measurement Conference}}. \bibinfo{pages}{73--88}.
\newblock


\bibitem[\protect\citeauthoryear{Mankins et~al\mbox{.}}{Mankins et~al\mbox{.}}{1995}]%
        {mankins1995technology}
\bibfield{author}{\bibinfo{person}{John~C Mankins} {et~al\mbox{.}}} \bibinfo{year}{1995}\natexlab{}.
\newblock \showarticletitle{Technology readiness levels}.
\newblock \bibinfo{journal}{{\em White Paper, April\/}} \bibinfo{volume}{6}, \bibinfo{number}{1995} (\bibinfo{year}{1995}), \bibinfo{pages}{1995}.
\newblock


\bibitem[\protect\citeauthoryear{Mirdita, Schulmann, Vogel, and Waidner}{Mirdita et~al\mbox{.}}{2024}]%
        {mirdita2023cure}
\bibfield{author}{\bibinfo{person}{Donika Mirdita}, \bibinfo{person}{Haya Schulmann}, \bibinfo{person}{Niklas Vogel}, {and} \bibinfo{person}{Michael Waidner}.} \bibinfo{year}{2024}\natexlab{}.
\newblock \showarticletitle{The CURE to vulnerabilities in RPKI validation}.
\newblock \bibinfo{journal}{{\em NDSS Symposium\/}} (\bibinfo{year}{2024}).
\newblock


\bibitem[\protect\citeauthoryear{Mirdita, Schulmann, and Waidner}{Mirdita et~al\mbox{.}}{2025}]%
        {dona:usenix2025}
\bibfield{author}{\bibinfo{person}{Donika Mirdita}, \bibinfo{person}{Haya Schulmann}, {and} \bibinfo{person}{Michael Waidner}.} \bibinfo{year}{2025}\natexlab{}.
\newblock \showarticletitle{{SoK: An Introspective Analysis of RPKI Security}}. In \bibinfo{booktitle}{{\em 34th {USENIX} Security Symposium}}. \bibinfo{publisher}{{USENIX} Association}.
\newblock


\bibitem[\protect\citeauthoryear{Mirdita, Shulman, and Waidner}{Mirdita et~al\mbox{.}}{2022}]%
        {mirdita2022poster}
\bibfield{author}{\bibinfo{person}{Donika Mirdita}, \bibinfo{person}{Haya Shulman}, {and} \bibinfo{person}{Michael Waidner}.} \bibinfo{year}{2022}\natexlab{}.
\newblock \showarticletitle{Poster: RPKI kill switch}. In \bibinfo{booktitle}{{\em Proceedings of the 2022 ACM SIGSAC Conference on Computer and Communications Security}}. \bibinfo{pages}{3423--3425}.
\newblock


\bibitem[\protect\citeauthoryear{NIST}{NIST}{2024}]%
        {nistmonitor}
\bibfield{author}{\bibinfo{person}{NIST}.} \bibinfo{year}{2024}\natexlab{}.
\newblock \bibinfo{booktitle}{{\em RPKI Monitor}}.
\newblock \bibinfo{type}{{T}echnical {R}eport}.
\newblock
\newblock
\shownote{"accessed 09/19/2024".}


\bibitem[\protect\citeauthoryear{of~Standards and Technology}{of~Standards and Technology}{2024}]%
        {xzutils}
\bibfield{author}{\bibinfo{person}{National~Institute of Standards} {and} \bibinfo{person}{Technology}.} \bibinfo{year}{2024}\natexlab{}.
\newblock \bibinfo{title}{{ CVE-2024-3094}}.
\newblock \bibinfo{howpublished}{\\\url{https://nvd.nist.gov/vuln/detail/CVE-2024-3094}}.   (\bibinfo{year}{2024}).
\newblock
\newblock
\shownote{Accessed 09/09/2024.}


\bibitem[\protect\citeauthoryear{of~the National Cyber~Director}{of~the National Cyber~Director}{2024}]%
        {whreport}
\bibfield{author}{\bibinfo{person}{Office of~the National Cyber~Director}.} \bibinfo{year}{2024}\natexlab{}.
\newblock \bibinfo{title}{{Roadmap to enhancing Internet Routing Security}}.
\newblock \bibinfo{howpublished}{\\\url{https://www.whitehouse.gov/wp-content/uploads/2024/09/Roadmap-to-Enhancing-Internet-Routing-Security.pdf}}.   (\bibinfo{year}{2024}).
\newblock


\bibitem[\protect\citeauthoryear{Snijders}{Snijders}{2024}]%
        {clientrelease}
\bibfield{author}{\bibinfo{person}{J Snijders}.} \bibinfo{year}{2024}\natexlab{}.
\newblock \bibinfo{booktitle}{{\em rpki-client release 9.1.}}
\newblock \bibinfo{type}{{T}echnical {R}eport}.
\newblock


\bibitem[\protect\citeauthoryear{Sun, Apostolaki, Birge{-}Lee, Vanbever, Rexford, Chiang, and Mittal}{Sun et~al\mbox{.}}{2021}]%
        {DBLP:journals/cacm/SunABVRCM21}
\bibfield{author}{\bibinfo{person}{Yixin Sun}, \bibinfo{person}{Maria Apostolaki}, \bibinfo{person}{Henry Birge{-}Lee}, \bibinfo{person}{Laurent Vanbever}, \bibinfo{person}{Jennifer Rexford}, \bibinfo{person}{Mung Chiang}, {and} \bibinfo{person}{Prateek Mittal}.} \bibinfo{year}{2021}\natexlab{}.
\newblock \showarticletitle{Securing internet applications from routing attacks}.
\newblock \bibinfo{journal}{{\em Commun. {ACM}\/}} \bibinfo{volume}{64}, \bibinfo{number}{6} (\bibinfo{year}{2021}), \bibinfo{pages}{86--96}.
\newblock


\bibitem[\protect\citeauthoryear{Tyson, Albert, and Brownsword}{Tyson et~al\mbox{.}}{2003}]%
        {tyson2003interpreting}
\bibfield{author}{\bibinfo{person}{Barbara Tyson}, \bibinfo{person}{Cecilia Albert}, {and} \bibinfo{person}{Lisa Brownsword}.} \bibinfo{year}{2003}\natexlab{}.
\newblock \bibinfo{booktitle}{{\em Interpreting Capability Maturity Model{\textregistered} Integration (CMMI{\textregistered}) for COTS-Based Systems}}.
\newblock \bibinfo{type}{{T}echnical {R}eport}. \bibinfo{institution}{Technical Report. Carnegie Mellon Software Engineering Institute}.
\newblock


\bibitem[\protect\citeauthoryear{Validator}{Validator}{2024a}]%
        {fortcve}
\bibfield{author}{\bibinfo{person}{Fort Validator}.} \bibinfo{year}{2024}\natexlab{a}.
\newblock \bibinfo{booktitle}{{\em Fort CVE List}}.
\newblock \bibinfo{type}{{T}echnical {R}eport}.
\newblock


\bibitem[\protect\citeauthoryear{Validator}{Validator}{2024b}]%
        {fortrelease}
\bibfield{author}{\bibinfo{person}{Fort Validator}.} \bibinfo{year}{2024}\natexlab{b}.
\newblock \bibinfo{booktitle}{{\em Fort Patch 1.6.3.}}
\newblock \bibinfo{type}{{T}echnical {R}eport}.
\newblock


\bibitem[\protect\citeauthoryear{Van~Hove, van~der Ham, and van Rijswijk-Deij}{Van~Hove et~al\mbox{.}}{2022}]%
        {van2022rpkiller}
\bibfield{author}{\bibinfo{person}{Koen Van~Hove}, \bibinfo{person}{Jeroen van~der Ham}, {and} \bibinfo{person}{Roland van Rijswijk-Deij}.} \bibinfo{year}{2022}\natexlab{}.
\newblock \showarticletitle{Rpkiller: Threat analysis from an RPKI relying party perspective}.
\newblock \bibinfo{journal}{{\em arXiv preprint arXiv:2203.00993\/}} (\bibinfo{year}{2022}).
\newblock


\bibitem[\protect\citeauthoryear{Vogel}{Vogel}{2024}]%
        {ripearticle}
\bibfield{author}{\bibinfo{person}{N Vogel}.} \bibinfo{year}{2024}\natexlab{}.
\newblock \bibinfo{booktitle}{{\em Crashing the Party: Vulnerabilities in RPKI Relying Party Software}}.
\newblock \bibinfo{type}{{T}echnical {R}eport}.
\newblock
\newblock
\shownote{"accessed 09/19/2024".}


\end{thebibliography}
